\documentclass[10pt, conference, compsocconf]{IEEEtran}
\IEEEoverridecommandlockouts  
\pdfoutput=1  

\usepackage{cite}

\usepackage[pdftex]{graphicx}

\usepackage{color}
\usepackage{listings}
\definecolor{darkgreen}{rgb}{0,0.5,0}
\definecolor{darkblue}{rgb}{0,0,0.5}
\definecolor{darkorange}{rgb}{0.5,0.25,0}
\usepackage{booktabs}
\usepackage{fixltx2e}

\usepackage{tikz}
\usepackage{textcomp}
\usepackage{hyperref}
\usepackage{lipsum}

\newcommand\copyrighttext{%
  \footnotesize \textcopyright 2016 IEEE. Personal use of this material is
  permitted.  Permission from IEEE must be
  obtained for all other uses, in any current or future media, including
  reprinting/republishing this material for advertising or promotional
  purposes, creating new
  collective works, for resale or redistribution to servers or lists, or reuse
  of any copyrighted component of this work in other works.
  DOI: \href{http://dx.doi.org/10.1109/IPDPSW.2016.108}{10.1109/IPDPSW.2016.108}}
\newcommand\copyrightnotice{%
\begin{tikzpicture}[remember picture,overlay]
\node[anchor=south,yshift=10pt] at (current page.south) {\fbox{\parbox{\dimexpr\textwidth-\fboxsep-\fboxrule\relax}{\copyrighttext}}};
\end{tikzpicture}%
}

\begin{document}

\title{Counting Triangles in Large Graphs on GPU}

\author{\IEEEauthorblockN{Adam Polak}
\IEEEauthorblockA{Department of Theoretical Computer Science\\
Faculty of Mathematics and Computer Science\\
Jagiellonian University in Krak\'ow\\
polak@tcs.uj.edu.pl}
}

\maketitle
\copyrightnotice

\begin{abstract}
The clustering coefficient and the transitivity ratio are concepts often used in
network analysis, which creates a need for fast practical algorithms for
counting triangles in large graphs. Previous research in this area focused on
sequential algorithms, MapReduce parallelization, and fast approximations.

In this paper we propose a parallel triangle counting algorithm for CUDA GPU.
We describe the implementation details necessary to achieve high performance and
present the experimental evaluation of our approach. The algorithm achieves
15 to 35 times speedup over our CPU implementation, and is capable
of finding 8.8 billion triangles in a 180 million edges graph in 12 seconds
on the Nvidia GeForce GTX 980 GPU.
\end{abstract}


\section{Introduction}

The number of triangles, i.e. cycles of length three, in an undirected graph
lays the foundation of the clustering coefficient and the transitivity
ratio -- concepts which are not only of theoretical interest
but are often applied to networks
analysis \cite{WATTS,EUBANK}. This creates a need for fast practical
algorithms capable of counting triangles in large graphs.

Previous research in this area focused mainly on sequential algorithms
\cite{SCHANK,LATAPY}, parallel algorithms for MapReduce model \cite{SURI},
and various approximation approaches \cite{DOULION,JHA}.

The emergence of frameworks for general-purpose computing on graphics
processor units (GPU), such as Nvidia CUDA, started a new branch of research in
parallel computing. General-purpose GPU offers the computing power of a small
cluster for much lower price, but it comes at a cost of certain
limitations imposed by the architecture of graphic cards. Single Instruction
Multiple Data model, high latency global memory, and small cache size
are obstacles particularly hard to overcome in case of memory-intensive and
highly irregular graph computations.

Nevertheless, a number of studies show that there are methods of dealing with
these issues and certain graph problems can be solved effectively on GPU --
examples include
minimum spanning tree \cite{VINEET},
connected components \cite{SOMAN},
breadth first search \cite{HONG,MERRILL},
and strongly connected components \cite{BARNAT}.

In this paper we propose a parallel triangle counting algorithm
together with its CUDA implementation, and discuss its performance.
Comparing to a single-threaded CPU solution, we achieve $8$ to $16$ times speedup
running on the Nvidia Tesla C2050 GPU, and $15$ to $35$ speedup 
running on the Nvidia GeForce GTX 980 GPU.
We also examine a setup with multiple GPUs.
We are able to further speed up the computation up to $2.8$ times when running
on four Tesla C2050 cards instead of one.

Similar studies have been conducted already by Leist et al.~\cite{LEIST},
and more recently by Chatterjee \cite{CHATTERJEE} and Green et al.~\cite{GREEN}.
As we argue later, our approach, despite being very simple, significantly 
outperforms the previous ones.

This paper is structured as follows.
Section \ref{sec:alg} outlines the algorithm we use.
Section \ref{sec:impl} provides the implementation details and discusses
the optimizations we employ.
Section \ref{sec:exp} carefully describes the experiments performed to evaluate
our implementation, and presents the results of these experiments.
Section \ref{sec:comp} compares our algorithm with other approaches to
the problem of counting triangles.
Section \ref{sec:concl} contains our conclusions.

\section{Algorithm}
\label{sec:alg}

\subsection{Known Sequential Algorithms}

Schank and Wagner \cite{SCHANK} present an extensive list of known sequential
algorithms for counting and listing triangles.
They analyze their theoretical time complexity,
and evaluate them experimentally on both synthetic and real world graphs. Two
algorithms, the \emph{edge-iterator} and \emph{forward}, appear to be the
winners of this comparison.
Their running times are $O(m\deg_{\max}$) and $O(m\sqrt{m})$,
respectively, where $m$ denotes the number of edges, and $\deg_{\max}$ denotes
the maximum degree of a vertex.
They perform similarly for graphs with low
deviation from the average degree, but the latter is more robust to skewed degree
distributions. Latapy \cite{LATAPY} simplifies the \emph{forward} algorithm,
reducing its memory
needs and running time, and makes its analysis more straightforward. This
modified version can be seen as a variant of the \emph{edge-iterator} algorithm
with an additional preprocessing phase. This allows a tighter bound on the
worst-case complexity and greatly reduces the amount of work to be done in the
subsequent counting phase, especially in the case of graphs with a high
degree deviation.

We choose \emph{forward} as a starting point for our parallel algorithm.
Both the preprocessing and
counting phases are not only easily parallelizable but they also rarely access
the memory in random fashion, which makes the algorithm a good fit for GPU.

\subsection{Sequential \emph{Forward} Algorithm}

Now we briefly describe the sequential \emph{forward} algorithm.
First, the algorithm sets
an arbitrary linear order $\prec$ on vertices which is consistent with their
degrees, i.e. $\deg(u) < \deg(v)$ implies $u \prec v$. Then, for every vertex
$u$, it filters its adjacency list removing vertices $v$ such that $v \prec u$.
Since $\prec$ is antisymmetric,
$u$ remains unremoved in the adjacency list of $v$.
This turns each undirected edge into a directed one, by choosing the orientation
from the vertex with the lower degree to the vertex with the higher degree.
In the final step of the preprocessing, the algorithm sorts adjacency lists
according to some arbitrary, previously fixed, linear order,
e.g. the natural order on vertices' numbers.

After this preprocessing
is done, the algorithm iterates over all edges and, for each edge, calculates
intersection of adjacency lists of both its ends. The total size of these
intersections is the number of triangles in the graph. Since adjacency lists are
sorted, each such intersection can be computed by a two pointers merge algorithm
(as in \emph{merge-sort})
in time linear in the sum of lengths of both lists.
Note that only edges that go forward according to $\prec$ are left.
Thus, every triangle is counted exactly once,
and it can be shown \cite{SCHANK,LATAPY} that no adjacency list is
longer than $\sqrt{m}$, which makes the whole algorithm run in $O(m\sqrt{m})$
time. 

\subsection{Parallel \emph{Forward} Algorithm}

The preprocessing phase is easily parallelized with a few \emph{prefix sum} and
\emph{sort} routines. We describe it in detail in the next section.

The counting phase is parallelized by assigning a single thread to each edge.
Each thread calculates adjacency lists intersection in sequential manner.
This gives us an $O(\sqrt{m})$ time algorithm running on $O(m)$ processors,
thus having the optimal work.

From a theoretical point of view, our algorithm does not have a polylogarithmic
time complexity, and thus does not put the problem into the class NC.
However, in practice, the speedup is always bounded by the number of available
processors (or cores in the case of GPU). Our algorithm has the optimal work and
$O(m)$ speedup, thus it is optimal for graphs with the number of edges greater
than the number of processors the algorithm is executed on.
This is generally the case since modern graphic cards usually have only a few
hundred up to few thousand cores.

\section{Implementation}
\label{sec:impl}

The implementation described in this section and all the tools used to run
the experiments described in the next section are available on
GitHub\footnote{\texttt{https://github.com/adampolak/triangles}}.

\subsection{Input Format}

Before going to implementation details it is important to note what the input
format of our algorithm is. It is often tempting to use a data representation that
is particularly suitable for a specific algorithm. However, in practice it is
rarely the case that the data is available in the selected format, and converting
it may take a significant amount of time.

In most works on graph computation on GPU the input format is either 
an \emph{adjacency list} \cite{HONG,BARNAT,MERRILL}
or an \emph{edge array} \cite{VINEET,SOMAN}. We decided to use the latter.
An \emph{edge array} is an array of structures, each structure composed of two
values -- identifiers of the two vertices a given edge connects.
We assume there are no self-loops nor
multiple edges, and each (undirected) edge appears exactly
twice, once in each direction.  We do not assume any particular order of the
edges.

The rationale for of our choice is similar to the one presented in \cite{SOMAN}.
An \emph{adjacency list} can be converted to an \emph{edge array} with a fast
and simple single-pass algorithm. The conversion in the opposite direction
requires sorting, which makes it much more computationally expensive.
Thus, using the \emph{edge array} representation makes the algorithm more
versatile in the sense that it can be used in various contexts without any
significant overhead for a format conversion.

The above intuitions can be further supported by the following numbers.
On the LiveJournal graph, which we use in the experiments
(for more details, see Section \ref{sec:exp}),
our CPU solution optimized for an \emph{adjacency list} input runs about $12$
seconds, while the solution optimized for an \emph{edge array} input is only
$2$ seconds slower. On the other hand, converting this graph from the
\emph{edge array} representation to the \emph{adjacency list} representation
takes about $7$ seconds.

\subsection{Preprocessing Phase}

The preprocessing phase consists of eight steps.
They make a heavy use of the Thrust library \cite{THRUST}.

\begin{enumerate}
  \item Copy the \emph{edge array} to the GPU memory.
  \item Calculate number of vertices using \emph{thrust::reduce} routine
    with \emph{thrust::maximum} operator, which computes the largest vertex
    identifier across both ends of all edges.
  \item Sort edges, according to the first vertex, in case of a tie according
    to the second vertex, using \emph{thrust::sort} routine, which performs
    radix sort.  This way the \emph{edge array} becomes a concatenated adjacency
    list of subsequent vertices, each list sorted.
  \item Calculate the \emph{node array}:
    $i$-th element of this array points to the first
    edge in the \emph{edge array} whose first vertex is $i$.
    This is done by running
    $m-1$ threads and letting $k$-th thread examine edges $k$ and $k+1$.
    If first vertices of these edges differ, the thread writes $k+1$ to an
    adequate cell of the \emph{node array}.
    It may happen that the thread stores this
    value in more than one cell when there is a vertex with an empty adjacency
    list.
  \item Mark edges going ``backwards'', i.e. from a vertex with the higher degree
    to a vertex with the lower degree.  In case of a tie, compare identifiers
    of vertices.  The degree of a vertex can be calculated quickly by subtracting
    two subsequent cells of the \emph{node array}.
  \item Remove ``backward'' edges using \emph{thrust::remove\_if} routine,
    which removes marked elements and compacts the \emph{edge array} preserving
    the order of elements that are not removed.
  \item Transform the \emph{edge array} from 
    an array of structures to a structure
    of arrays. We call this step \emph{unzipping}.
  \item Recalculate the \emph{node array}.
\end{enumerate}

\subsection{Counting Phase}

Triangles are counted with the following kernel performing the two pointers
merge algorithm. The $i$-th thread deals with edges whose index in the
\emph{edge array} modulo the number of threads equals $i$. For each such edge
the thread sequentially computes the size of the intersection of the
neighborhoods of the both ends of this edge.

\begin{lstlisting}
__global__ void CountTriangles(
    int m,
    const int* __restrict__ edge,
    const int* __restrict__ node,
    uint64_t* result) {
  int start = blockDim.x * blockIdx.x + threadIdx.x;
  int step = gridDim.x * blockDim.x;
  uint64_t count = 0;

  // for each assigned edge (u, v)
  for (int i = start; i < m; i += step) {
    int u = edge[i], v = edge[m + i];
    int u_it = node[u];
    int u_end = node[u + 1];
    int v_it = node[v];
    int v_end = node[v + 1];

    // run two pointers merge algorithm
    int a = edge[u_it], b = edge[v_it];
    while (u_it < u_end && v_it < v_end) {
      int d = a - b;
      if (d <= 0) a = edge[++u_it];
      if (d >= 0) b = edge[++v_it];
      if (d == 0) ++count;
    }
  }

  result[start] = count;
}
\end{lstlisting}

After the kernel is done, elements of \emph{result} array are summed, using
\emph{thrust::reduce} routine, to obtain the total triangle count, and
the algorithm terminates.

As usual, a careful choice of the number of threads and blocks to run is crucial
to achieve high performance. 
We tuned these parameters experimentally, using a grid search approach,
with the number of threads per block going through powers of two from $32$ to
$1024$ and the number of blocks per multiprocessor varying from $1$ to $16$
in steps of $1$. We concluded that it is optimal to run $64$ threads per block
and $8$ blocks per each multiprocessor.
These numbers are optimal, or nearly optimal, for all the graphs we used in our
experiments, as well as for all the three devices we had for our tests,
i.e. NVS 5200M, Tesla C2050, and GTX 980.
It is worth noting that on GTX 980 a similar performance can be achieved with
other combinations giving $512$ threads per multiprocessor, e.g. $256$ threads
per block and $2$ blocks per multiprocessor. However, this is not the case
for the two older devices.

\subsection{Optimizations}

\subsubsection{Unzipping Edges.}
\emph{CountTriangles} kernel runs $13\%$ to $32\%$ faster when the 
\emph{edge array} is given as a structure of arrays.
Conversion from an array of structures to a structure of arrays
is very fast, i.e. it takes less than $30$ milliseconds for all the graphs
we used in our experiments, largest of them having more than $200$ million edges.

\subsubsection{Sorting Edges as 64-bit Integers.}
Sorting edges with \emph{thrust::sort} is approximately $5$ times faster when
the \emph{edge array} is passed to it as an array of 64-bit integers instead of
an array of pairs of 32-bit integers.
When using this optimization it is important to remember that,
because of the endianness, it produces a slightly different
ordering -- edges are ordered by the second vertex then, in case of a tie, by the first.

\subsubsection{Avoiding Unnecessary Reads.}
Compare our preliminary version of a while loop in \emph{CountTriangles} kernel:
\begin{lstlisting}
while (u_it < u_end && v_it < v_end) {                                       
  int d = edge[u_it] - edge[v_it];                           
  if (d <= 0) ++u_it;                                                        
  if (d >= 0) ++v_it;                                                        
  if (d == 0) ++count;                                                       
}     
\end{lstlisting}
with our final version:
\begin{lstlisting}
int a = edge[u_it], b = edge[v_it];
while (u_it < u_end && v_it < v_end) {
  int d = a - b;
  if (d <= 0) a = edge[++u_it];
  if (d >= 0) b = edge[++v_it];
  if (d == 0) ++count;
}
\end{lstlisting}

The preliminary version reads two values form the \emph{edge array} in every loop
execution, while the final version reads only one value when no triangle
is found. This difference seems unimportant because these
unnecessarily read values are often cached, and the preliminary version
performs less work in divergent branches. Nevertheless, the final version
runs $36\%$ to $48\%$ faster.

\subsubsection{Read-Only Data Cache.}
Starting with the Nvidia Kepler architecture the L1 cache is disabled for global
memory. However, read-only data can be cached in the texture cache.
Since our algorithm relies heavily on the \emph{edge array} caching,
it is crucial that the \emph{edge array} is marked with the
\texttt{const \_\_restrict\_\_} qualifiers, which allow compiler to use the
texture cache. This change results in $17\%$ to $66\%$ faster kernel
execution on the Kepler and Maxwell architectures.

\subsubsection{Reducing Warp Size.}
We can simulate reducing the warp size by doubling the number of threads
and making half of the threads within a warp idle.
Although our final implementation does not benefit from this method, we find
it worth noting since it allowed $30\%$ faster kernel execution at earlier
stages of the kernel's development.
We believe it is due to the fact that, when a read misses
the cache and a thread has to wait for the global memory,
all other threads in the warp have to wait as well.
Reducing the warp size makes less threads affected by a single cache miss.
This effect is especially significant for our algorithm because cache
misses happen at different moments for different threads.

\subsubsection{CPU Preprocessing for Very Large Graphs.}
\label{sec:lowmemory}
Sorting in the step 3 of the preprocessing phase
requires the largest amount of the global memory of the GPU.
If the input graph is too large to fit into the memory in this step,
we use a slightly modified version of the preprocessing.
First, we use the CPU to calculate vertex degrees and remove backward
edges. It runs slower than on the GPU but halves the input size.
Then, we can move to GPU to sort and \emph{unzip} edges
and calculate the \emph{node array}.
This optimization allows to process graphs twice larger than without it.

\subsubsection{Unsuccessful Optimization Attempts.}
We tried the virtual warp-centric method \cite{HONG}, collaborative
reading to shared memory, and assembler level prefetching.
None of these optimizations increased the performance of our implementation,
probably due to a high overhead compared to possible gains.

\subsection{Multi-GPU Setup}

We propose a simple extension of our algorithm to a multi-GPU setup.
The preprocessing phase is run just on a single GPU, then the
\emph{edge array} and \emph{node array} are copied to the remaining devices,
and each device iterates over its allotted subset of edges.

The speedup of this approach is limited by the Amdahl's law. The preprocessing
time is roughly proportional to the number of edges,
while the counting phase time appears to be proportional to
the number of triangles. The fraction of the execution time spent on the
preprocessing varies depending on the graph -- for graphs that we use in our
experiments this fraction ranges from $0.08$ to $0.76$,
which gives the maximum speedup for $4$ GPUs between $3.23$ and $1.22$.
This is roughly consistent with our experimental results.
The biggest speedups are obtained for Kronecker graphs, which have
large triangles to edges ratios.

A less trivial approach to the multi-GPU parallelization would probably require
splitting the graph into (not necessarily disjoint) subgraphs, which then
can be processed independently \cite{CHU,SURI}. However, it is not clear if the
obtained speedup would compensate the overhead caused by the splitting phase.
We do not cover this issue in this paper, but we find it a viable direction
for future research.

\section{Experiments}
\label{sec:exp}

To evaluate the performance of our implementation we ran it on a number of
graphs and compared its running time with
a baseline single-threaded CPU implementation. The
baseline implementation is our own implementation of the \emph{forward}
algorithm, and it is slightly faster than the one provided in~\cite{LATAPY}.

The graphs we used are:
Internet topology graph, LiveJournal online social network, and Orkut online
social network from Stanford Large Network Dataset Collection \cite{SNAP};
Citeseer and DBLP co-paper networks and Kronecker R-MAT graphs
from 10th DIMACS Implementation Challenge \cite{DIMACS}; Barab\'asi-Albert
network \cite{BARABASI}; Watts-Strogatz network \cite{WATTS}.
Table \ref{table} summarizes basic properties of these graphs.

\begin{table*}[!t]
\small
\renewcommand{\arraystretch}{1.2}
\caption{Experimental results.}
\label{table}
\centering

\begin{tabular}{@{}l rrr r rr rr rr@{}}
\toprule
Graph & Nodes & Edges & Triangles & CPU & \multicolumn{2}{c}{Tesla C2050} & \multicolumn{2}{c}{$4$ $\times$ Tesla C2050} & \multicolumn{2}{c}{GTX 980}\\
 & & & & Time [ms] & Time [ms] & Speedup & Time [ms] & Speedup & Time [ms] & Speedup \\ 
\midrule
Real world graphs &&&&&&&&\\
\quad  Internet topology & $1.7$M   & $22$M & $ 29$M &$  3459$&$  277$&$12.49$&$  306$&$0.91$&$  186$&$18.60$\\
\quad  LifeJournal       & $  4$M   & $69$M & $178$M &$ 13829$&$  951$&$14.54$&$  947$&$1.00$&$  540$&$25.61$\\
\quad  Orkut             & $3.1$M   &$234$M & $628$M &$ 82558$&$^\dagger9690$&$ ^\dagger8.52$&$ ^\dagger7580$&$^\dagger1.28$&$ 2815$&$29.33$\\
\quad  Citeseer          & $0.4$M   & $32$M & $872$M &$  4990$&$  578$&$ 8.63$&$  456$&$1.27$&$  329$&$15.17$\\
\quad  DBLP              & $0.5$M   & $30$M & $442$M &$  4712$&$  446$&$10.57$&$  410$&$1.09$&$  239$&$19.72$\\
Synthetic graphs &&&&&&&&\\
\quad  Kronecker 16      & $2^{16}$ &  $5$M &$ 119$M &$  2810$&  $179$&$15.70$&$   97$&$1.85$&$   82$&$34.27$\\
\quad  Kronecker 17      & $2^{17}$ & $10$M &$ 288$M &$  6957$&  $476$&$14.62$&$  219$&$2.17$&$  219$&$31.77$\\
\quad  Kronecker 18      & $2^{18}$ & $21$M &$ 688$M &$ 17808$& $1274$&$13.98$&$  499$&$2.55$&$  558$&$31.91$\\
\quad  Kronecker 19      & $2^{19}$ & $44$M &$1626$M &$ 45947$& $3434$&$13.38$&$ 1304$&$2.63$&$ 1443$&$31.84$\\
\quad  Kronecker 20      & $2^{20}$ & $89$M &$3804$M &$116811$& $9308$&$12.55$&$ 3296$&$2.82$&$ 3942$&$29.63$\\
\quad  Kronecker 21      & $2^{21}$ &$182$M &$8816$M &$297426$&$^\dagger33150$&$ ^\dagger8.97$&$^\dagger13624$&$^\dagger2.43$&$12009$&$24.77$\\
\quad  Barab\'asi-Albert & $0.2$M   & $20$M &$   3$M &$  5508$&  $327$&$16.84$&$  263$&$1.24$&$  155$&$35.54$\\
\quad  Watts-Strogatz    & $1$M     & $50$M &$ 219$M &$  9627$&  $589$&$16.34$&$  576$&$1.02$&$  324$&$29.71$\\
\bottomrule
\end{tabular}
\end{table*}

Experiments were run on the Nvidia Tesla C2050 GPU, Nvidia GeForce GTX 980 GPU,
and Intel Xeon X5650 CPU.
We compiled the binaries using the NVCC 7.0 and G++ 4.8 compilers with the
\texttt{-O3} optimization level.

We measured wall clock time. We started each measurement just before the \emph{edge
array} is copied from CPU to GPU, and finished it right after the final result was
copied from GPU to CPU and the GPU memory was freed. Before the measurement we
called \emph{cudaFree(NULL)} to preinitialize CUDA context, because otherwise
the first call to \emph{cudaMalloc} takes additional $100$ milliseconds.

Each experiment was run five times. In the paper we report the mean values.
The standard deviation never exceeded $0.05$ of the mean.

\begin{figure}
  \centering
  \includegraphics[width=0.5\textwidth]{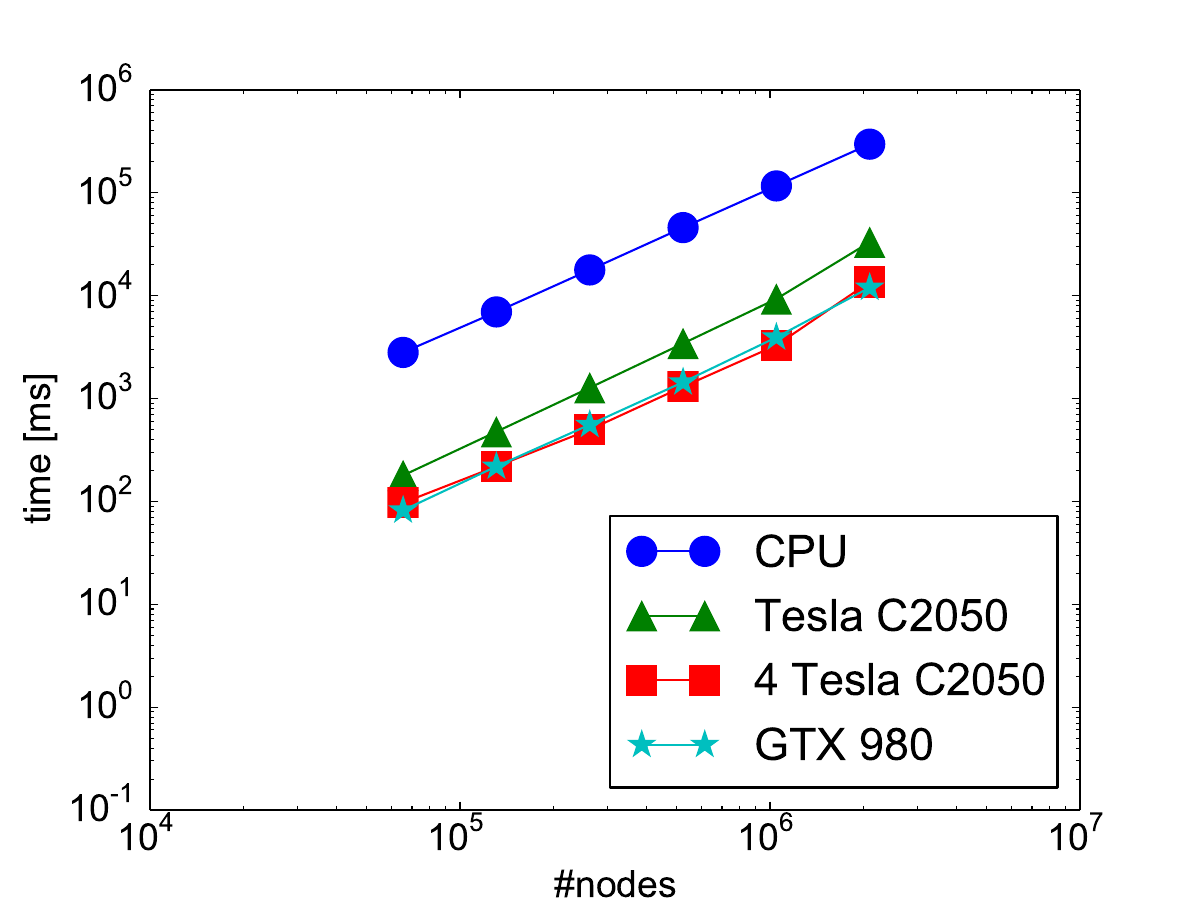}
  \caption{Experimental results for synthetic Kronecker R-MAT graphs.}
  \label{fig:kron}
\end{figure}

The results are presented in Table~\ref{table}.
All execution times are given in milliseconds.
The first and the third speedup columns show the GPU over CPU speedup, while
the second speedup column shows the 4 GPU over 1 GPU speedup.
Measurements marked with $\dagger$ denote graphs too large to fit into the GPU
memory, which required part of the preprocessing phase to be run on the CPU
(as described in Section \ref{sec:lowmemory}).
It accounts for lower performance in case of these graphs.
The results for Kronecker graphs are also presented visually in
Figure~\ref{fig:kron}.

In order to asses the efficiency of our implementation of the
\emph{CountTriangles} kernel we used profiler to measure
the cache hit rate and memory bandwidth during kernel execution
on the GTX 980 card. These measurements are presented in Table~\ref{tab:prof}.
We consider the cache hit in the $75\%$ to $80\%$ range being a good result.
The GTX 980 card offers $224$ GB/s of peak memory bandwidth.
Our implementation achieves about half of this value, which is also satisfactory.
In conclusion, the profiling analysis shows that there is a room for
improvement, but it is not large.

\begin{table}
\small
\renewcommand{\arraystretch}{1.2}
\caption{Profiling results on GTX 980.}
\label{tab:prof}
\centering
\begin{tabular}{@{}l rr@{}}
\toprule
Graph & Cache hit rate & Bandwidth [GB/s] \\
\midrule
Real world graphs &&\\
\quad  Internet topology & $80.78\%$ & $ 95.90$ \\
\quad  LifeJournal       & $79.73\%$ & $100.28$ \\
\quad  Orkut             & $82.71\%$ & $ 98.55$ \\
\quad  Citeseer          & $76.68\%$ & $117.92$ \\
\quad  DBLP              & $78.14\%$ & $112.96$ \\
Synthetic graphs &&\\
\quad  Kronecker 16      & $80.95\%$ & $143.99$ \\
\quad  Kronecker 17      & $79.75\%$ & $134.33$ \\
\quad  Kronecker 18      & $78.35\%$ & $128.33$ \\
\quad  Kronecker 19      & $77.59\%$ & $122.60$ \\
\quad  Kronecker 20      & $76.78\%$ & $113.37$ \\
\quad  Kronecker 21      & $75.81\%$ & $ 93.65$ \\
\quad  Barab\'asi-Albert & $64.45\%$ & $137.56$ \\
\quad  Watts-Strogatz    & $74.55\%$ & $116.82$ \\
\bottomrule
\end{tabular}
\end{table}

\section{Comparison to Related Work}
\label{sec:comp}

In the past there were various attempts to count triangles faster than with
sequential algorithms. MapReduce approach to the problem\cite{SURI}
has significant overhead, and even for moderately sized graphs the execution
time is in the order of minutes. It is beneficial to use it for extremely large
graphs, with the number of edges in the order of one billion.
Another approach is to use an heuristic approximation algorithm
\cite{DOULION,JHA}. Such algorithms provide good speedups and usually need
little memory, but it comes at the cost of getting only an approximate
triangle count, which can differ from the actual count usually by a few percent.

Our CUDA implementation of the parallel triangle counting algorithm,
presented in this paper,
achieves $8$ to $16$ times speedup over our optimized single-threaded CPU solution,
when run on the Nvidia Tesla C2050 GPU, and $15$ to $35$ speedup 
running on the Nvidia GeForce GTX 980 GPU.
By using four graphic cards instead of one we can further speedup computation
up to $2.8$ times, especially when the number of triangles in the input graph 
is much larger than the number of edges.

Taking into account the speedups achieved by using GPU to solve other
graph problems -- e.g.
$50$ times for minimum spanning tree \cite{VINEET},
$10$ times for connected components \cite{SOMAN},
$4$ to $30$ times for breadth first search \cite{MERRILL},
and $5$ to $40$ times for strongly connected components \cite{BARNAT} --
our results seem satisfactory.

The first work we can try to directly compare to is \cite{LEIST}.
Unfortunately, such comparison is not easy for a number of reasons.
First, the source code for their approach is not available, so we can
compare only to the numbers provided in the paper. Second, the paper solves
a slightly different problem, which is computing the clustering coefficient.
It requires computing the number of triangles but also the number of two-edge
paths in the input graph. Fortunately, the latter part is not harder than the
former, so we can assume this gives our algorithm at most two times advantage.
Third, in the cited paper, experiments were run on the Nvidia GeForce GTX 480
graphic card, which is very similar to the Nvidia Tesla C2050. Nevertheless,
they are different devices. Lastly, the execution times measured during
these experiments are presented only on plots, so we can obtain only approximate
numerical values.
Having said that, our algorithm is $45$ times faster than the previous one on
the Barab\'asi-Albert network, and $7$ times faster on the Watts-Strogatz
network (the numbers change to $20$ and $3$, respectively, when comparing
runtimes in the four cards setups).
We believe this difference is largely due to our choice of the \emph{forward}
algorithm.

Another GPU algorithm, proposed in \cite{CHATTERJEE}, is evaluated on the
Nvidia Tesla C1060 which is a slower device than those we used.
However, the author reports running times in the order of $20$ seconds for
graphs with $2000$ nodes, which means our approach is orders of magnitude
faster.

The most recent work on the topic \cite{GREEN} proposes much more elaborate
algorithm, in which also the adjacency list intersection step is parallelized.
The algorithm was evaluated on a number of real world graphs, two of which
(Citeseer and DBLP) also appear in our experiments. The evaluation was performed
on the Nvidia Tesla K40, which is not less powerful than the Nvidia Tesla C2050,
which we used. Despite this, our algorithm achieves roughly two times lower
execution times for these two graphs.

Yet another approach to count triangles faster is to use a multi-core CPU.
According to \cite{LEIST}, a parallel counting algorithm running on a
$6$-core CPU with $12$ virtual hyper-threading cores can achieve $7$ times
speedup over a single-threaded solution. Assuming this result scales well
with increasing number of CPUs, it should be possible to achieve performance
similar to what we present, on a multiprocessor machine.
However, both price and energy consumption of such a setup are likely to be
higher than in our case.

\section{Conclusions and Future Work}
\label{sec:concl}

In this paper we proposed a parallel triangle counting algorithm for CUDA.
We proved it can be implemented efficiently, and described details necessary to do so.
The algorithm achieves $15$ to $35$ times speedup over our baseline single-threaded
solution, and is capable of finding $8.8$ billion triangles in a $180$ million
edges graph in $12$ seconds on the Nvidia GeForce GTX 980 GPU.
Despite being very simple, our algorithm significantly outperforms,
to our best knowledge, all triangle counting algorithms for GPU up to date.

We plan on extending our research in two directions, which we now briefly
discuss.

First, it would be interesting to check if methods from \cite{CHU,SURI} can be
applied, without a too big overhead,
to split the graph into subgraphs which can be
processed independently. This could give a better multi-GPU solution, and what
is more important, this would allow to count triangles in graphs which do not
fit into the GPU memory -- which is one of the biggest limitations of our current
algorithm.

Second, it might be beneficial to use a different counting algorithm for a small
subset of vertices with largest degrees. A natural candidate for such algorithm
is matrix multiplication \cite{AYZ}.

\section*{Acknowledgment}
This research was supported by the Polish Ministry of Science and Higher
Education program ``Diamentowy Grant''.

\bibliographystyle{IEEEtran}
\bibliography{IEEEabrv,triangles}

\end{document}